# Comment on the paper "*Constitutive boundary conditions and paradoxes in nonlocal elastic nanobeams. International Journal of Mechanical Sciences, 121, 2017, 151-156*" by Giovanni Romano, Raffaele Barretta, Marina Diaco, and Francesco Marotti de Sciarra

**Mohamed Shaat***

*Mechanical Engineering Department, Zagazig University, Zagazig 44511, Egypt*

Romano et al. [2017a], in the considered study, discussed the different constitutive models proposed by Eringen according to their abilities to give well-posed nonlocal boundary value problems that admit exact solutions. They concluded:

- A nonlocal field problem formed based on Eringen's constitutive model admits a unique solution only if the nonlocal field fulfils constitutive boundary conditions (formed in Romano et al. [2017a]).

- The paradoxes of the nonlocal results of previous studies stem from the incompatibility between the constitutive boundary conditions and equilibrium conditions imposed on the nonlocal field.

- The fully nonlocal elasticity model secretes ill-posed nonlocal field problems, which admit no solutions.

- The local/nonlocal mixture elasticity renders well-posed nonlocal field problems, but only if the local component is a strictly positive fraction of the mixture. The problem will anyway tend to ill-posedness when this fraction tends to vanish.

Here, we comment against these arguments and demonstrate that:

- No exact solution is exist for any nonlocal boundary value problem formed by any of Eringen's nonlocal constitutive laws.

- The nonlocal boundary value problem formed by Eringen's model admits no exact solution even if the constitutive boundary conditions defined by Romano et al. [2017a] are fulfilled.

- No physical meaning of the constitutive boundary conditions defined by Romano et al. [2017a] since the original problem has no solution.

-----------------------------------------------------------------------------------------------

∗ *E-mail addresses:* shaatscience@yahoo.com; shaat@nmsu.edu (M. Shaat).

**Demonstration 1: Non-existence of exact solutions for nonlocal boundary value problems formed by Eringen's nonlocal model**

The constitutive law of Euler-Bernoulli beams can be written in the general form [Romano et al., 2017a]:

$$M_\lambda(x) = \int_a^b \phi_\lambda(x - y) . (K . \chi_{el})(y) \, dy \tag{1}$$

where $\phi_\lambda(x)$ is an attenuation function. $\chi_{el}$ is the beam curvature.

This nonlocal constitutive law may depend on a strongly singular kernel [Polizzotto, 2001]:

$$\phi_\lambda(x) = \frac{1}{2L_c} \exp\left(-\frac{|x|}{L_c}\right) \tag{2}$$

which gives the "fully nonlocal" constitutive law with the form:

$$M_\lambda(x) = \frac{1}{2L_c} \int_a^b \exp\left(-\frac{|x - y|}{L_c}\right) . (K . \chi_{el})(y) \, dy \tag{3}$$

An alternative kernel, which is weakly singular, can be considered [Polizzotto, 2001]:

$$\phi_\lambda(x) = \xi\delta(x) + \frac{1 - \xi}{2L_c} \exp\left(-\frac{|x|}{L_c}\right) \tag{4}$$

where $\delta(x)$ is the Dirac delta function.

The substitution of the kernel (4) into (1) gives the "local/nonlocal mixture" constitutive law [Polizzotto, 2001]:

$$M_\lambda(x) = \xi(K . \chi_{el})(x) + \frac{(1 - \xi)}{2L_c} \int_a^b \exp\left(-\frac{|x - y|}{L_c}\right) . (K . \chi_{el})(y) \, dy \tag{5}$$

It is evident that the local/nonlocal mixture constitutive law is a special case of the fully nonlocal constitutive law [Polizzotto, 2001]. The local/nonlocal mixture constitutive law depends on a weakly singular kernel, and $\xi$ is a parameter that defines the degree of the singularity of the nonlocal kernel. When $\xi = 0$, the nonlocal kernel is strongly singular and the fully nonlocal constitutive law is reproduced. The degree of the singularity decreases with an increase in $\xi$ value. The singularity is totally eliminated only when $\xi = 1$ where the constitutive law reproduces the constitutive law of the local beam. It follows that the fully nonlocal and the local/nonlocal mixture constitutive laws represent the same nonlocal constitutive model but with different degrees of singularity.

The equivalent differential local/nonlocal mixture constitutive model has the form [Romano et al. 2017a]:

$$M_\lambda(x) - L_c^2 M_\lambda''(x) = (K . \chi_{el})(x) - L_c^2 \xi(K . \chi_{el})''(x) \tag{6}$$

with the constitutive boundary conditions [Romano et al. 2017a]:

$$M_\lambda'(0) - \frac{1}{L_c}M_\lambda(0) = \xi(K.\chi_{el})'(0) - \frac{\xi}{L_c}(K.\chi_{el})(0)$$

$$M_\lambda'(L) + \frac{1}{L_c}M_\lambda(L) = \xi(K.\chi_{el})'(L) + \frac{\xi}{L_c}(K.\chi_{el})(L)$$

(7)

The differential fully nonlocal constitutive model is recovered when $\xi = 0$:

$$M_\lambda(x) - L_c^2 M_\lambda''(x) = (K.\chi_{el})(x)$$

(8)

with the constitutive boundary conditions:

$$M_\lambda'(0) - \frac{1}{L_c}M_\lambda(0) = 0$$

$$M_\lambda'(L) + \frac{1}{L_c}M_\lambda(L) = 0$$

(9)

Romano et al. [2017a] set a condition to obtain a well-posed nonlocal boundary value problem that admits a solution. According to Romano et al. [2017a], a solution of a nonlocal field problem exists **only if** the constitutive boundary conditions (7) (for the local/nonlocal mixture model) or (9) (for the fully nonlocal model) are fulfilled by the formed bending field, and no contradictions are exist between the equilibrium conditions and these constitutive boundary conditions. Thus, if these constitutive conditions are not fulfilled, the nonlocal field problem admits no solution at all.

Here, we demonstrate that a nonlocal boundary value problem formed by any one of Eringen's constitutive laws admits no exact solution, even if the constitutive boundary conditions defined by Romano et al. [2017a] are fulfilled. Accordingly, we reveal that the ill-posedness of Eringen's nonlocal constitutive laws is an inherited trait of the nonlocal model, and it cannot be eliminated.

To demonstrate this, let us consider the equilibrium differential equation for the elastostatic of Euler-Bernoulli beams modeled using Eringen's model:

$$M_\lambda''(x) = q(x)$$

(10)

where $q(x)$ is the applied transverse load. $M_\lambda(x)$ is the bending field. A general solution of equation (10) can be written in the form:

$$M_\lambda(x) = M_h + M_p \quad \text{with} \quad M_h = C_1 x + C_2$$

(11)

where $C_1$ and $C_2$ are constants to be determined using the equilibrium boundary conditions. $M_h$ and $M_p$ are homogeneous and particular solutions, respectively. According to equations (10) and (11), the nonlocal bending field has the same form as the local bending field (*this is the main defect of Eringen's model*). The bending field (11) is then substituted into a constitutive law ((3) or (5)), which should be solved for the beam curvature $\chi_{el}(x)$. In fact, neither the fully nonlocal constitutive law (3) nor the local/nonlocal mixture constitutive law (5) admit any exact solution for $\chi_{el}(x)$ for a given bending field obtained from the equilibrium equation (10). To demonstrate this, the bending field (11) is substituted into (3):

$$M_h(x) = C_1 x + C2 = -M_p(x) + \xi(K.\chi_{el})(x) + \frac{(1-\xi)}{2L_c} \int_a^b \exp\left(-\frac{|x-y|}{L_c}\right).(K.\chi_{el})(y)\, dy \qquad (12)$$

and into (5):

$$M_h(x) = C_1 x + C2 = -M_p(x) + \frac{1}{2L_c} \int_a^b \exp\left(-\frac{|x-y|}{L_c}\right).(K.\chi_{el})(y)\, dy \qquad (13)$$

It is clear that there are no exact solutions for equations (12) and (13). The right-hand sides of these equations are nonlinear for any value of the curvature $\chi_{el}$ while the left-hand sides (the homogeneous part of the bending field) are linear. There is only one case at which exact solutions of these constitutive models exist, which depends on the particular part of the bending moment ($M_p$) and the equilibrium boundary conditions to be specially formed depending on the nonlocal character of the considered beam. However, no practical application can be observed for the latter case of the loading and equilibrium boundary conditions.

Next, it is demonstrated that the right-hand sides of equations (12) and (13) produce two exponential terms, which, obviously, form a nonlinear expression. These two exponential terms make the existence of an exact solution of the beam curvature is not possible, as it will be discussed later.

**Demonstration 2: The constitutive boundary conditions have no physical sense**

What about the constitutive boundary conditions (7) and (9)? The Proposition 3.1 in [Romano et al., 2017a] states "*The constitutive integral equation Eq.([5]) with the special kernel Eq.(3) admits, for any $\lambda > 0$, either a unique solution or no solution at all, depending on whether or not the bending field fulfils the constitutive boundary conditions.*" Thus, Romano et al. [2017a] tied the existence of a solution of the nonlocal field problem to the fulfilment of the constitutive boundary conditions. However, there is no physical meaning of the constitutive boundary conditions (equations (7) and (9)) since the original problem has no solution. Thus, even if these constitutive boundary conditions are fulfilled, a solution still is not exist.

For the demonstration and without loss of generality, let us consider a simple case of a beam subjected to point force ($\mathcal{F}$) and moment ($\mathcal{M}$), *i.e.* $M_p = 0$ and $M_\lambda(x) = M_h(x) = \mathcal{F}(L-x) + \mathcal{M}$. Consequently, the local/nonlocal mixture constitutive model (equations (6) and (7)) can be written as follows:

$$(K.\chi_{el})(x) - L_c^2 \xi(K.\chi_{el})''(x) - \mathcal{F}(L-x) - \mathcal{M} = 0 \qquad (14)$$

with the constitutive boundary conditions:

$$-\mathcal{F} - \frac{1}{L_c}(\mathcal{F}L + \mathcal{M}) = \xi(K.\chi_{el})'(0) - \frac{\xi}{L_c}(K.\chi_{el})(0) \qquad (15)$$

$$-\mathcal{F} + \frac{1}{L_c}\mathcal{M} = \xi(K.\chi_{el})'(L) + \frac{\xi}{L_c}(K.\chi_{el})(L)$$

For a constant beam stiffness, $K$, a general solution of equation (14) can be obtained as follows:

$$\chi_{el}(x) = d_1\exp\left(\frac{x}{L_c\sqrt{\xi}}\right) + d_2\exp\left(-\frac{x}{L_c\sqrt{\xi}}\right) + \frac{\mathcal{F}(L-x) + \mathcal{M}}{K} \tag{16}$$

where $d_1$ and $d_2$ are two constants that shall be determined by applying the constitutive boundary conditions (equation (15)). Thus, the constitutive boundary conditions of the local/nonlocal mixture model are fulfilled automatically. It should be noted that the local/nonlocal mixture constitutive law gives the curvature depending on two exponential terms, as presented in equation (16). Later it will be demonstrated that these two exponential terms lead to an unavoidable singularity of the curvature field and its derivatives.

To examine the validity of the obtained solution from the differential local/nonlocal mixture constitutive model, which satisfied the constitutive boundary conditions (equation (15)), it is substituted into the integral constitutive law (5), which gives:

$$M_\lambda(x) = \xi K\left(d_1\exp\left(\frac{x}{L_c\sqrt{\xi}}\right) + d_2\exp\left(-\frac{x}{L_c\sqrt{\xi}}\right) + \frac{\mathcal{F}(L-x) + \mathcal{M}}{K}\right)$$

$$+ K\frac{(1-\xi)}{2L_c}\int_a^b\exp\left(-\frac{|x-y|}{L_c}\right)\cdot\left(d_1\exp\left(\frac{y}{L_c\sqrt{\xi}}\right)\right.$$

$$\left.+ d_2\exp\left(-\frac{y}{L_c\sqrt{\xi}}\right) + \frac{\mathcal{F}(L-x) + \mathcal{M}}{K}\right)dy \tag{17}$$

By performing the integration, the bending field from the local/nonlocal mixture constitutive law is obtained as follows:

$$M_\lambda(x) = a_1\exp\left(\frac{x}{L_c}\right) + a_2\exp\left(-\frac{x}{L_c}\right) + \mathcal{F}(L-x) + \mathcal{M} \tag{18}$$

where

$$a_1 = \frac{\exp\left(-\frac{L(\sqrt{\xi}+1)}{L_c\sqrt{\xi}}\right)}{2}\left(Kd_1(\xi + \sqrt{\xi})\exp\left(\frac{2L}{L_c\sqrt{\xi}}\right)\right.$$

$$\left.+ (-(L + Lc)\mathcal{F} + \mathcal{F}L + \mathcal{M})(\xi - 1)\exp\left(\frac{L}{L_c\sqrt{\xi}}\right) + Kd_2(\xi - \sqrt{\xi})\right) \tag{19}$$

$$a_2 = \frac{(\mathcal{F}L_c + \mathcal{F}L + \mathcal{M})(\xi - 1) + K\xi(d_1 + d_2) - K\sqrt{\xi}(d_1 - d_2)}{2}$$

It is clear that the bending moment $M_\lambda(x)$ obtained from the nonlocal constitutive law is nonlinear due to two exponential terms in equation (18). These two exponential terms demonstrate an inconsistency between bending fields obtained from the equilibrium equation and bending fields obtained from the nonlocal constitutive law (the bending field obtained from the equilibrium equation is linear and equals $\mathcal{F}(L-x) + \mathcal{M}$). Thus, the local/nonlocal mixture constitutive law cannot reproduce the bending field obtained from the equilibrium equation; instead, it provides two residual exponential terms, as shown in equation (18). Because of this inconsistency, the nonlocal boundary value problem is ill-posed when formed using the local/nonlocal mixture constitutive law, contrary to the claimed well-posedness exposed by Romano et al. [2017a].

It is interesting to observe that the nonlocal constitutive law secretes two exponential terms in each of the curvature function ($\chi_{el}(x)$) and the moment function ($M_\lambda(x)$), as presented in equations (16) and (18). These exponential terms form nonlinear expressions of the beam curvature and moment with unavoidable singularities of the curvature and moment fields and their derivatives. For the fully nonlocal constitutive law, Romano et al. [2017a] derived an equation similar to equation (18) representing the bending field (see equation (31) in [Romano et al., 2017a]). They demonstrated that these residual exponential terms lead to unavoidable singularity of the bending moment, shear force, and loading field. The singularity is a verification of the non-existence of a solution of the nonlocal boundary value problem [Romano et al., 2017a; 2017b].

It follows from the previous discussion that the nonlocal boundary value problem formed using any of Eringen's nonlocal constitutive laws suffers of (1) an inconsistency between the equilibrium equation and the nonlocal constitutive law (2) and a singularity of the nonlocal fields. Given these observations, the question about the possibility of the total elimination of the inconsistency revealed in equation (18) and overcoming the singularity of the nonlocal fields now arises. It is possible to eliminate the inconsistency revealed in equation (18). However, the singularity of the nonlocal fields is unavoidable because it is not possible to eliminate all the exponential terms revealed in equations (16) and (18) (*these exponential terms are the source of the singularity*).

Now, let us examine the possibility of eliminating the inconsistency revealed in equation (18). This inconsistency can be eliminated by choosing values of $d_1$ and $d_2$ that make the coefficients $a_1$ and $a_2$ in equation (18) zeros for any value of the parameters $K, \mathcal{F}, \mathcal{M}, \xi, L,$ and $L_c$. Therefore, $d_1$ and $d_2$ are obtained for zero-coefficients $a_1$ and $a_2$, as follows:

$$d_1 = \frac{\left(-(-\mathcal{F}L_c + \mathcal{M})(\xi-1)^2 \exp\left(\frac{L}{L_c\sqrt{\xi}}\right) + \left(\mathcal{F}(L_c+L) + \mathcal{M}\right)\left(\xi^2 + 2\sqrt{\xi} - 2\xi^{3/2} - 1\right)\right)}{K\left(\xi - \sqrt{\xi}\right)\left(\left(\xi + 2\sqrt{\xi} + 1\right)\exp\left(\frac{2L}{L_c\sqrt{\xi}}\right) - \xi + 2\sqrt{\xi} - 1\right)}$$

$$d_2 = \frac{\exp\left(\frac{L}{L_c\sqrt{\xi}}\right)\left(-\left(\xi^2 - \xi - \sqrt{\xi} + \xi^{3/2}\right)(\mathcal{F}(L_c+L) + \mathcal{M})\exp\left(\frac{L}{L_c\sqrt{\xi}}\right) + \left((-\mathcal{F}L_c + \mathcal{M})\left(\xi^2 - \xi + \sqrt{\xi} - \xi^{3/2}\right)\right)\right)}{K\xi\left(\left(\xi + 2\sqrt{\xi} + 1\right)\exp\left(\frac{2L}{L_c\sqrt{\xi}}\right) - \xi + 2\sqrt{\xi} - 1\right)}$$

(20)

Although the inconsistency is eliminated when $d_1$ and $d_2$ are defined, as presented in equation (20), the singularity of the nonlocal fields still exists. To demonstrate this fact, the beam curvature $\chi_{el}(x)$ and the local-type loading field $P(x) = K \frac{d^2 \chi_{el}(x)}{dx^2}$ are plotted for different values of $L_c$ and $\xi$ parameters, as shown in Figures 1 and 2. It is evident from the figures that the considered nonlocal fields exhibit singular behaviors at the beam's ends. Values of these nonlocal fields sharply increase at the ends (tend to infinity).

The intensity of the singular behavior increases due to successive differentiations. Thus, the singularity of the loading field $P(x) = K \frac{d^2 \chi_{el}(x)}{dx^2}$ is sharper than that of the beam curvature $\chi_{el}(x)$. Moreover, it follows from Figures 1 and 2 that the degree of the singularity decreases with an increase in $L_c$ and/or $\xi$ parameters. These singularities demonstrate the non-existence of a solution of a nonlocal problem formed using any of Eringen's constitutive laws.

The existence of singularities can be proved by considering the solution of the local beam. A consistent nonlocal model reproduces the solution of a local beam when $\xi = 1$ and/or $L_c = 0$. The local solutions of the considered cantilever beams are $\chi(x) = -x + 1$ (for $\mathcal{F} = 1$, $\mathcal{M} = 0$, and $K = 1$) and $\chi(x) = 1$ (for $\mathcal{F} = 0$, $\mathcal{M} = 1$, and $K = 1$) and the loading field shall be completely vanished (*i.e.* $P(x) = 0$). It follows from Figures 1 and 2 that Eringen's constitutive models secrete inconsistent results with unavoidable singularities. For example, when $\xi = 0.999$, non-vanishing loading field $P(x)$ is obtained exhibiting singularity at the beam ends. The singularity is sharp when $L_c = 0.001$, and it decreases as $L_c$ value increases. A zero loading field is expected only when $\xi > 0$ and $L_c \to \infty$. This is inconsistent with the fact that a local solution is obtained when $\xi \to 1$ and/or $L_c \to 0$. This demonstrates the ill-posedness of nonlocal problems formed using Eringen's constitutive laws.

The unavoidable singularity of the nonlocal fields can be analytically demonstrated from the loading field function $P(x)$, which can be obtained from equation (16), as follows:

$$P(x) = K \frac{d^2 \chi_{el}(x)}{dx^2} = \frac{K}{Lc^2 \xi} \left( d_1 \exp\left( \frac{x}{L_c \sqrt{\xi}} \right) + d_2 \exp\left( -\frac{x}{L_c \sqrt{\xi}} \right) \right) \qquad (21)$$

Equation (21) reveals a non-vanishing loading field for all values of the $L_c$ and $\xi$ parameters. This field is completely vanished only when $L_c \to \infty$ and $\xi > 0$.

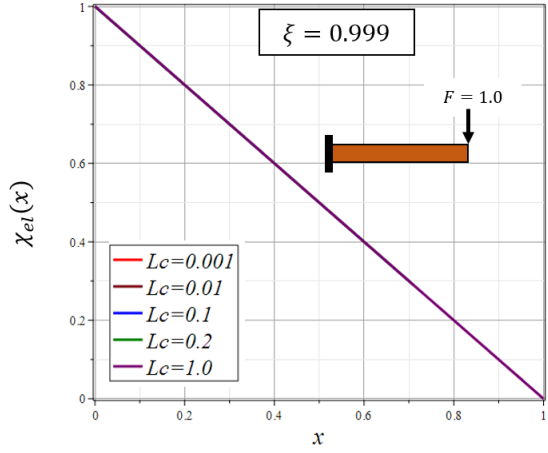

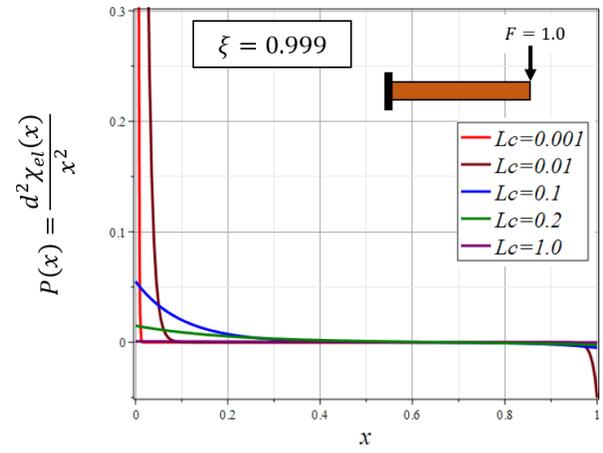

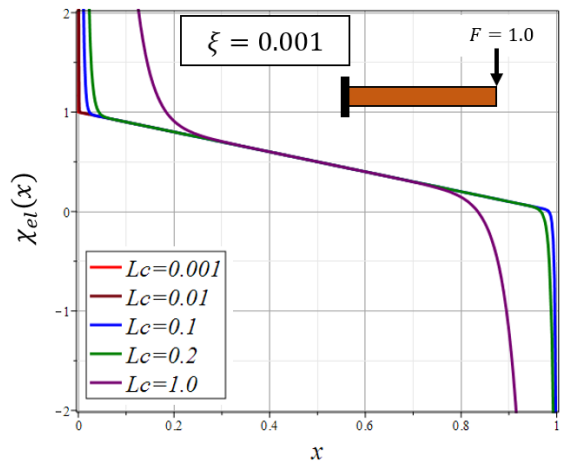

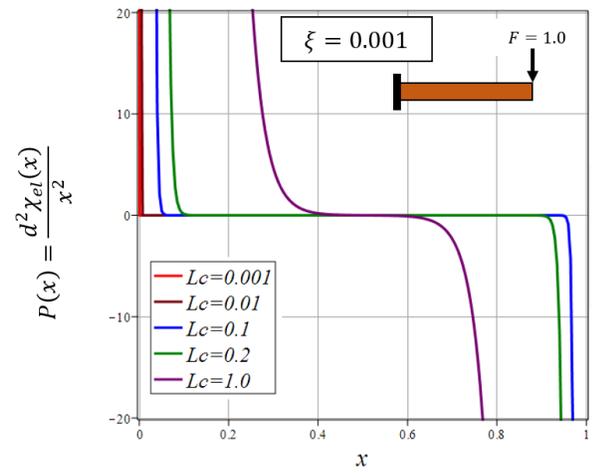

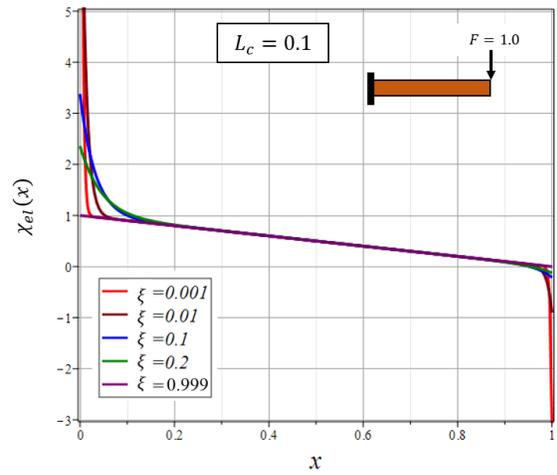

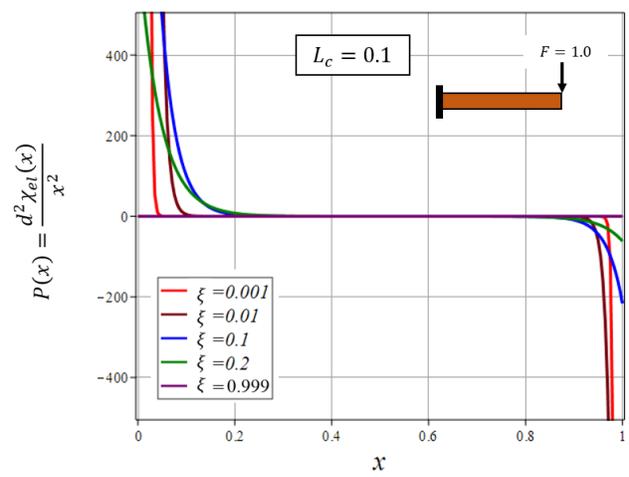

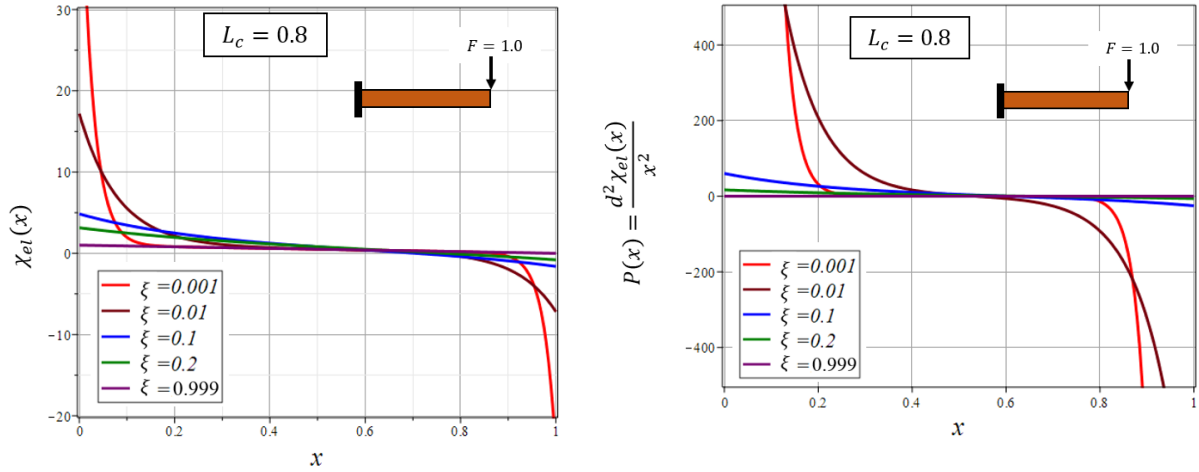

Figure 1: Singular behaviors of the elastic curvature, $\chi_{el}(x)$, and the local-type loading field $P(x) = K \dfrac{d^2\chi_{el}(x)}{dx^2}$ of a cantilever beam subjected to a point force at its free end ($K = 1$).

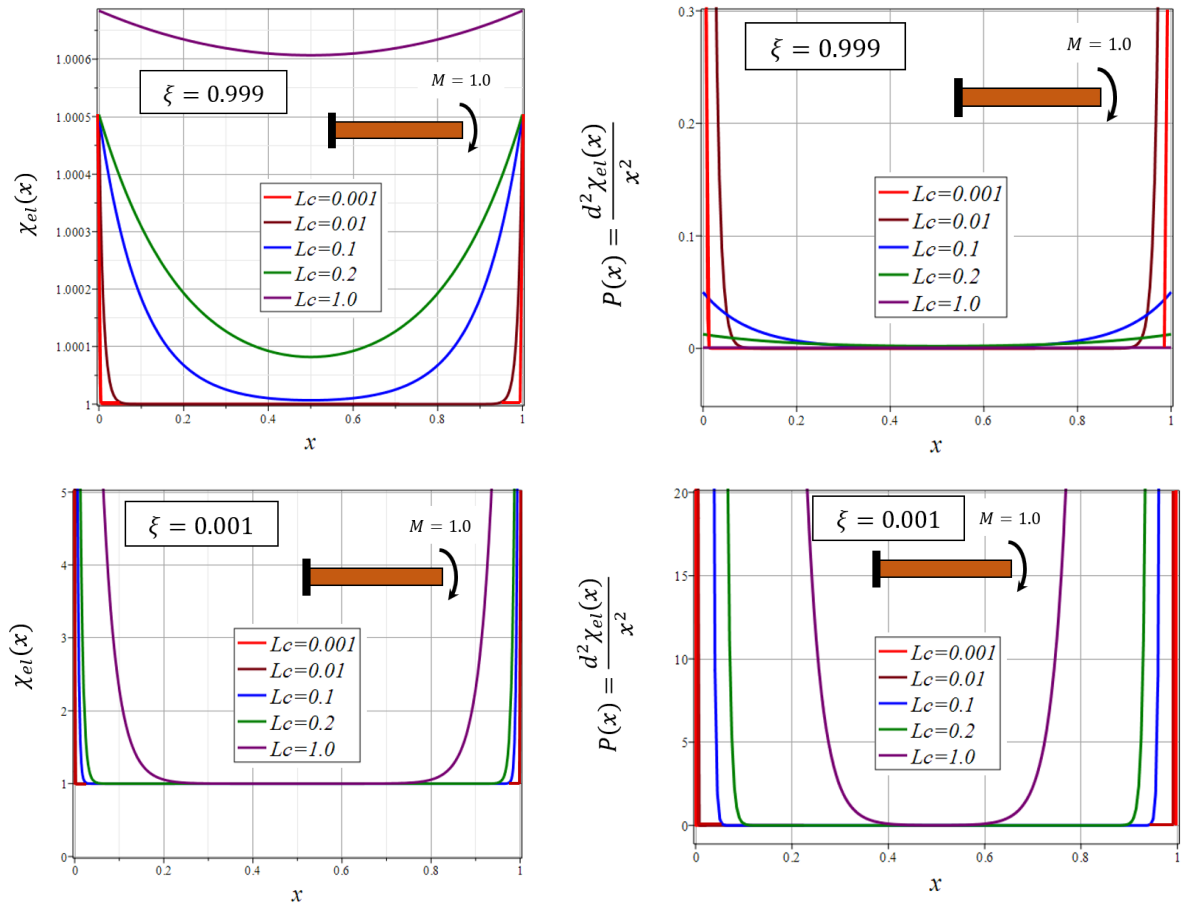

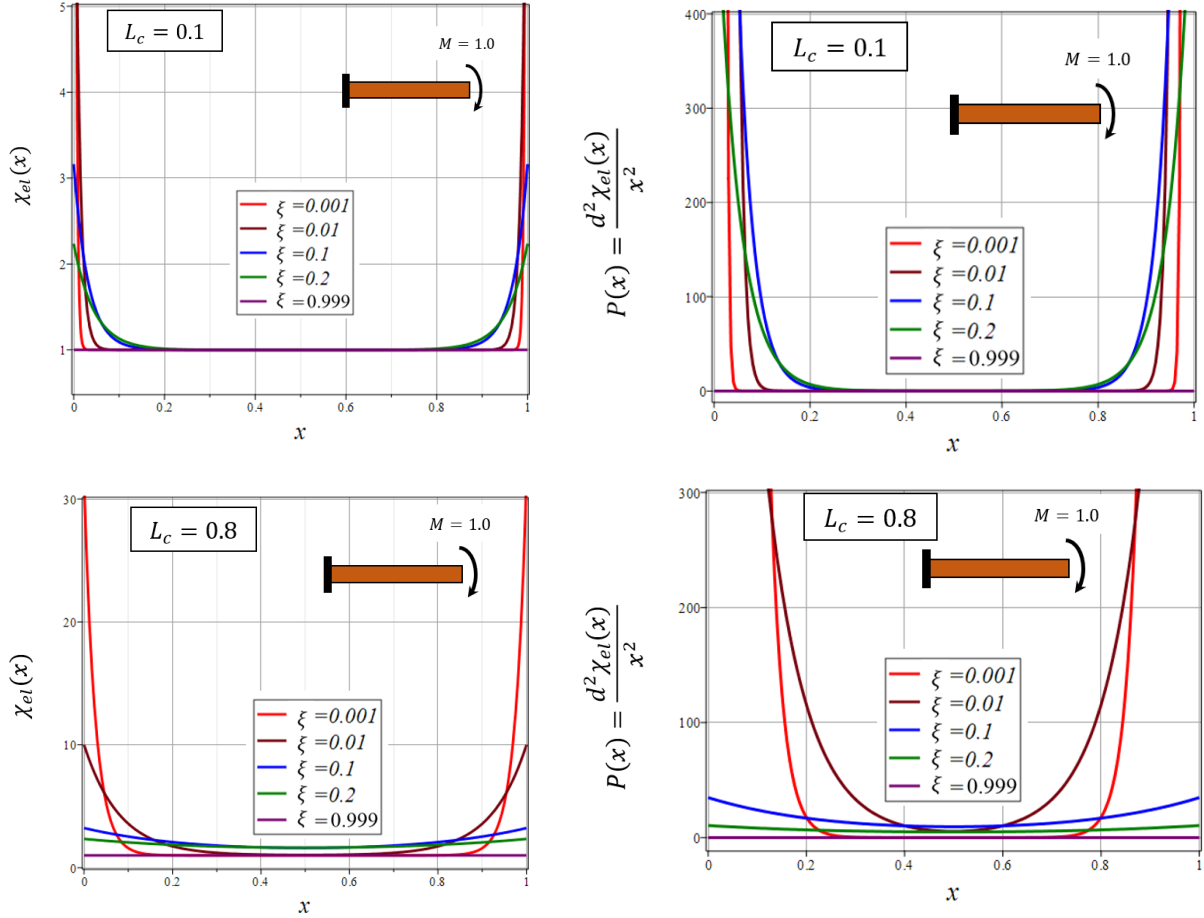

Figure 2: Singular behaviors of the elastic curvature, $\chi_{el}(x)$, and the local-type loading field $P(x) = K\frac{d^2\chi_{el}(x)}{dx^2}$ of a cantilever beam subjected to a moment at its free end ($K = 1$).

The previous demonstrations reveal the non-existence of exact solutions of nonlocal boundary value problems formed by Eringen's constitutive models is due to an inconsistency between the equilibrium equation and the nonlocal constitutive law and an unavoidable singularity of the nonlocal fields. The non-existence of the solutions of nonlocal problems demonstrated here demolishes the claimed Proposition 3.1 in [Romano et al., 2017a].

## Conclusions

No exact solution is exist for any nonlocal boundary value problem formed by any of Eringen's nonlocal constitutive models. The local/nonlocal mixture constitutive law does not secrete well-posed nonlocal problems as claimed by Romano et al. [2017a]. However, it is, like other nonlocal constitutive models, secretes ill-posed nonlocal problems that give no exact solutions at all. Moreover, the condition of the fulfilment of the constitutive boundary conditions proposed by Romano et al. [2017a] has no physical meaning because the original problem has no exact solution. The non-existence of exact solutions of nonlocal boundary value problems formed by Eringen's constitutive models is due to an inconsistency between the equilibrium equation and the nonlocal constitutive law and an unavoidable singularity of the nonlocal fields. In conclusion, the ill-posedness of Eringen's model cannot be eliminated where the homogeneous parts of the nonlocal fields cannot be properly formed using the nonlocal constitutive model.